\documentclass{PoS}
\usepackage{color}
\usepackage{amsmath}
\usepackage{amsfonts}
\usepackage{amssymb}
\usepackage{tabularx}
\usepackage{amsthm}
\usepackage{epsf}
\usepackage{booktabs}
\usepackage{graphicx}

\title{Flavored tetraquark spectroscopy}

\ShortTitle{Flavored tetraquark spectroscopy}

\author{\speaker{Andrea L. Guerrieri}\\
        Dipartimento di Fisica and INFN, Università di Roma 'Tor Vergata'\\
        Via della Ricerca Scientifica 1, I-00133 Roma, Italy\\
        E-mail: \email{andrea.guerrieri@roma2.infn.it}}
\author{Mauro Papinutto, Alessandro Pilloni, Antonio D. Polosa\\
	Dipartimento di Fisica and INFN, 'Sapienza' Università di Roma\\
	P.le Aldo Moro 5, I-00185 Roma, Italy}
\author{Nazario Tantalo\\
	CERN, PH-TH, Geneva, Switzerland and\\
	Dipartimento di Fisica and INFN, Università di Roma 'Tor Vergata' \\
	Via della Ricerca Scientifica 1, I-00133 Roma, Italy}

\abstract{The recent confirmation of the charged charmonium like resonance Z(4430) by the LHCb experiment
strongly suggests the existence of QCD multi quarks bound states.
Some preliminary results about hypothetical flavored tetraquark mesons are reported.
Such states are particularly amenable to Lattice QCD studies as their interpolating operators
do not overlap with those of ordinary hidden-charm mesons.}

\FullConference{The 32nd International Symposium on Lattice Field Theory,\\
		23-28 June, 2014\\
		Columbia University New York, NY}

\begin{document}

\section{Introduction}

The recent confirmation of the charged resonant state $Z(4430)$ by LHCb~\cite{Aaij:2014jqa} 
strongly suggests the existence of genuine compact tetraquark mesons in the QCD spectrum.
Among the many phenomenological models, it seems that only
the diquark-antidiquark model in its type-$II$ version can accomodate in a unified description the 
puzzling spectrum of the exotics~\cite{Maiani:2014aja}.
Although diquark-antidiquark model has success in describing the observed exotic spectrum, it also predicts 
a number of unobserved exotic partners.
Recently, it has proposed a mechanism à la Feshbach to explain the experimental lack of 
those states~\cite{Papinutto:2013uya,Guerrieri:2014gfa}.
Lattice QCD could provide useful insights on the nature of exotic mesons, 
though the numerical situation is still unclear due to both theoretical and tecnical reasons~\cite{Prelovsek:2013cra, Alexandrou:2012rm}.

\section{Flavored tetraquarks}

Our study is based on the proposal reported in~\cite{Esposito:2013fma}, where a theoretical framework for Lattice QCD numerical studies of 
"pure tetraquark" states is setup and 
possible phenomenological consequences are proposed in order to reveal these hypothetical hadrons in current experiments.
From a field theory point of view the definition of the exotic states, as the $J^{PC}=1^{++}$ $X(3872)$ or the charged $J^{PG}=1^{++}$ $Z(3900)$, it's a challenging task. 
The reason is that these states, having ordinary flavor quantum numbers, can be created from the vacuum by mesonic interpolating operators.
On the other hand, if one allows exotic flavor structures, as shown in~\cite{Esposito:2013fma}, it is possible to give a field theory meaningful definition 
of a tetraquark interpolating operator.
In this perspective, we will focus on operators with flavor content
\begin{equation}
\left[ c c\right] \left[ \bar q_1 \bar q_2 \right] \hspace{10mm} \bar q_1, \bar q_2 = \bar u, \bar d,
\end{equation}
the notation $[ q_1 q_2 ]$ means that quarks $q_1$ and $q_2$ form a diquark.
Operators of this kind have four valence quarks, being the valence number $N_{val}$ defined as
\begin{equation}
N_{val}=\sum_f |q_f|, \hspace{10mm} Q_f = \int d^3 x \bar \psi_f \gamma^0 \psi_f, \hspace{10mm} Q_f \left | \psi \right\rangle=q_f \left | \psi \right\rangle.
\end{equation}
Due to the open flavor structure, it's impossible to correlate quarks belonging to the same source.
This excludes also the presence of disconnected diagrams contributing to their correlation functions  
making simpler their numerical evaluation .
The quantum numbers of the charmed diquark is fixed by symmetry to
\begin{equation}
 [cc] = \left| \bar{3}_c(A), ~\ J^{P}\!=\!1^+(S)\right\rangle,
\end{equation}
where by $(S)$ and $(A)$ we indicate the symmetry/antisymmetry of a configuration.
The choices left for the light-light diquark are
\begin{eqnarray}
\left[ \bar{q}_1 \bar{q}_2 \right]_G &= \left| 3_c(A), \,1_I(A) ,\,J^{P}\!=\!0^+(A)\right\rangle\\
\left[\bar{q}_1 \bar{q}_2 \right]_B &= \left| 3_c(A), \, 3_I(S),\, J^{P}\!=\!1^+(S)\right\rangle.
\end{eqnarray}
The resulting diquark-antidiquark spectrum is shown in Table~\ref{tab:states}.
\begin{table}[h]
\centering
\begin{tabular}{c|c}
\hline
\multicolumn{2}{c}{$\mathcal{T}$ states}\\
\hline
``Good'', $J^P=1^+$, I=0 & ``Bad'', $J^P=0^+,1^+,2^+$, I=1\\
\hline
$\mathcal{T}^+\; ([cc]{[\bar u\bar d\,]_A})$ & $\mathcal{T}^0\; ([cc][\bar u\bar u])$\\
& $\mathcal{T}^{++}\; ([cc][\bar d\bar d])$\\
& $\mathcal{T}^+\; ([cc]{[\bar u\bar d\,]_S})$\\
\hline
\end{tabular}
\caption{$\mathcal{T}$ spectrum in the $J^P=1^+$ channel as predicted by diquark antidiquark model. Denomination GOOD or BAD comes from the quark model, where GOOD diquarks are expected to be lighter then BAD ones. Notice the possible existence of a 
doubly charged state $\mathcal{T}^{++}$.}
\label{tab:states}
\end{table}
We notice that, among the predicted states, there is a doubly-charged particle $\mathcal{T}^{++}$ that, if discovered, should confirm the validity of the model in contrast with the molecular
paradigm as
a one expects that a weakly interacting molecular state would fall apart due to Coulomb repulsion preventing the formation of whatever resonant structure.

\section{Simulation strategy}
In our analysis we consider a basis of five interpolating operators in the $J^{P}=1^+$, $I=0$ sector 
and a basis of three operators for the $I=1$ sector with same spin and parity. 
The interpolating operators chose for the $I=0$ sector are listed in Eqs.~\ref{O1}-\ref{O5}.
\begin{eqnarray}
\mathcal{O}_1=\varepsilon^{ijk}\varepsilon^{lmk}\bar{c}^i_c(x)\gamma^Ac^j(x) ~\ (\bar{u}^l(x)\gamma^5d_c^m(x)-\bar{d}(x)^l\gamma^5u_c^m(x)) \hspace{5mm} good \,\mathcal{T^+} 
\label{O1}\\
\mathcal{O}_{2}=\bar{u}(x)\gamma^Ac(x) \, \bar{d}(x)\gamma^5 c(x) - \bar{d}(x)\gamma^Ac(x) \, \bar{u}(x)\gamma^5 c(x) \hspace{5mm} D^0D^{*+}-D^{*0}D^+\label{O2}\\
\mathcal{O}_{3}=\bar{u}\gamma^Ac \left[ \vec{p}=\vec{0} \right] \, \bar{d}\gamma^5 c - \bar{d}\gamma^Ac \left[ \vec{p}=\vec{0}\right]\, \bar{u}\gamma^5 c \hspace{5mm} 
D^0D^{*+}-D^{*0}D^+\label{O3}\\
\mathcal{O}_{4}=\varepsilon^{ABC} \bar{u}(x)\gamma^Bc(x) \, \bar{d}(x)\gamma^C c(x) \hspace{5mm} D^{*0}D^{*+}\label{O4}\\
\mathcal{O}_{5}=\varepsilon^{ABC} \bar{u}\gamma^Bc \left[ \vec{p}=\vec{0} \right] \, \bar{d}\gamma^C c \hspace{5mm}D^{*0}D^{*+}\label{O5}
\end{eqnarray}
During the numerical analysis we invert propagators both on point like and stochastic sources. In particular, the correlation functions involving operators as those in Eqs.~\ref{O1}, \ref{O2}, \ref{O4} need point like inverted propagators, while those for~\ref{O3}, \ref{O5} stochastic propagators. 
The notation $[\vec{p}=\vec{0}]$ in Eqs.~\ref{O3}, \ref{O5} means that the two bilinears forming the operator are integrated separately. These are expected to have a large overlap, 
respectively, with the states of $DD^*$ and $D^*D^*$ mesons at rest\footnote{In our simulations the $D^*$ meson is stable because the simulated pion mass and the physical size of the lattice volume forbid its decay into a $D$ $\pi$ final state.}.
The operator~\ref{O2}, \ref{O4} are local products of quark bilinears and, in contrast with the preceding operators, we expect they can interpolate well also states with 
moving back to back mesons. 
We have implemented a tetraquark operator, Eq.~\ref{O1}, in order to spot a possible exotic state.
Analogous considerations are valid for the operators used in the $I=1$ sector, listed in Eqs.~\ref{O11}-\ref{eq:I1}.
\begin{eqnarray}
\bar{\mathcal{O}}_1=\varepsilon^{ijk}\varepsilon^{lmk}\bar{c}^i_c(x)\gamma^Ac^j(x) ~\ (\bar{u}^l(x)\gamma^Bd_c^m(x)+\bar{d}^l(x)\gamma^Bu_c^m(x)) \varepsilon^{ABC} \hspace{5mm} bad \,\mathcal{T^+} 
\label{O11}\\
\bar{\mathcal{O}}_{2}=\bar{u}(x)\gamma^Ac(x) ~\ \bar{d}(x)\gamma^5 c(x) + \bar{d}(x)\gamma^Ac(x) ~\ \bar{u}(x)\gamma^5 c(x) \hspace{5mm} D^0D^{*+}+D^{*0}D^+
\label{O12}\\
\bar{\mathcal{O}}_{3}=\bar{u}\gamma^Ac\left[ \vec{p}=\vec{0}\right] \, \bar{d}\gamma^5 c + \bar{d}\gamma^Ac \left[ \vec{p}=\vec{0}\right]\, \bar{u}\gamma^5 c \hspace{5mm} D^0D^{*+}+D^{*0}D^+.
\label{eq:I1}
\end{eqnarray}
The excited energy levels are extracted using the generalized eigenvalue problem (GEP)~\cite{Luscher:1990ck}.
We have computed all the entries of the correlation function matrix $C_{ij}=\langle 0 | \mathcal{O}_i (t) \mathcal{O}_j^\dag (0) | 0 \rangle$ for $i,j=1,\dots,N_{op}$, 
with $N_{op}$ the number of operators in the basis, and solved the problem
\begin{equation}
C(t)\psi = \lambda(t,t_0)C(t_0) \psi.
\end{equation}
The initial time $t_0$ is tuned in order to have the maximum numerical precision in determining the eigenvalues.
The resulting eigenvalues, labeled by an index $n$, decay exponentially with the $n$-th energy level up to exponentially suppressed deviations
\begin{equation}
\lambda_{n}(t,t_0) \sim e^{-E_{n}(t-t_0)}.
\end{equation} 
As a final remark, each operator is doubled using a gaussian smearing on fermion fields
\begin{equation}
\hat{S}=\frac{1+\alpha \Delta}{1 + 6\alpha},
\end{equation}
where $\Delta$ is the Laplacian over the spatial coordinates.
The number of smearing steps chosen is $50$ and $\alpha=0.5$.
\section{Numerical results}
\begin{figure}[htbp]
\begin{center}
\begin{minipage}[l]{.40\textwidth}
\centering
\includegraphics[scale=0.65]{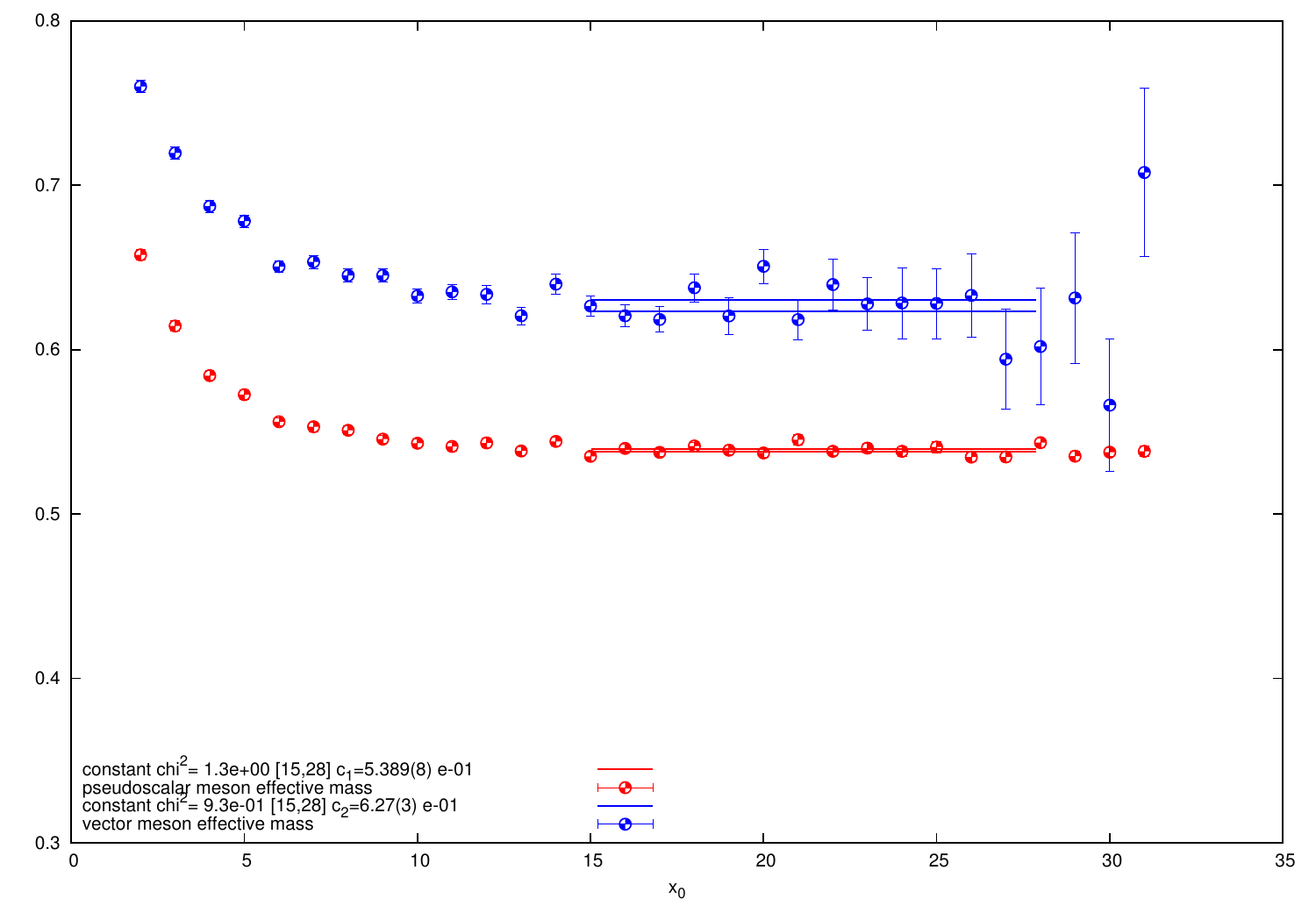}
\end{minipage}
\hspace{40mm}
\begin{minipage}[c]{.30\textwidth}
\centering
\begin{tabular}{c|c}
\hline
	 & thresholds in\\
	 & lattice units\\
\hline
\small{$DD^*$} & $1.166(4)$\\
\small{$D(1)D^*(-1)$} & $1.230(4)$\\
\small{$D^*D^*$} & $1.254(7)$\\
\small{$D^*(1)D^*(-1)$} & $1.314(6)$\\
\hline
\end{tabular}
\end{minipage}
\end{center}
\caption{In Figure are shown the ground states respectively of the $0^{-+}$ heavy light sector (red points) and of the $1^{--}$ sector(blue points).
The horizontal bars denote the value of the effective mass within one $\sigma$. In table are reported the possible free mesons $D$ $D^{(*)}$ energy levels.}
\label{fig:soglie}
\end{figure}

\begin{figure}[htbp]
\begin{center}
\begin{minipage}[l]{.40\textwidth}
\centering
\includegraphics[scale=0.6]{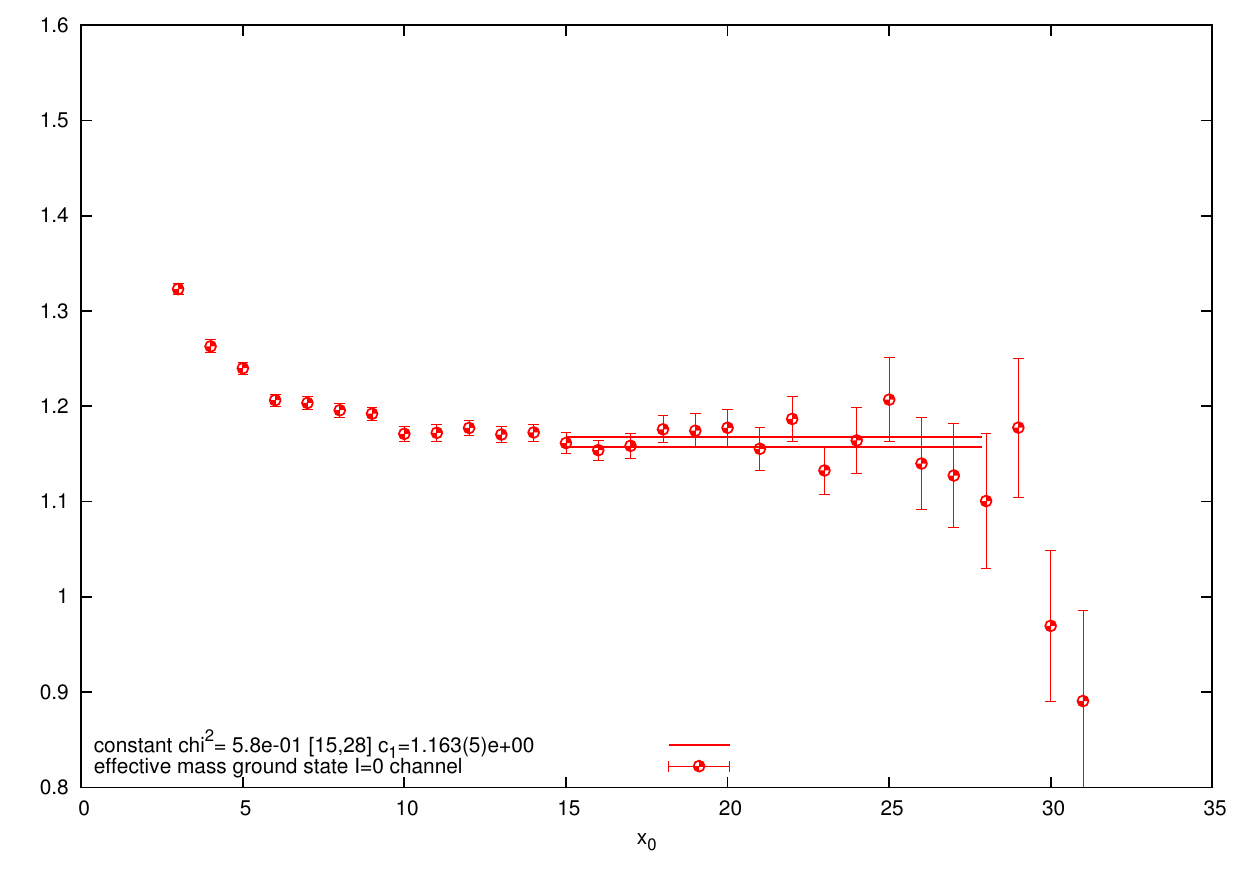}
\end{minipage}
\hspace{10mm}
\begin{minipage}[c]{.40\textwidth}
\centering
\includegraphics[scale=0.6]{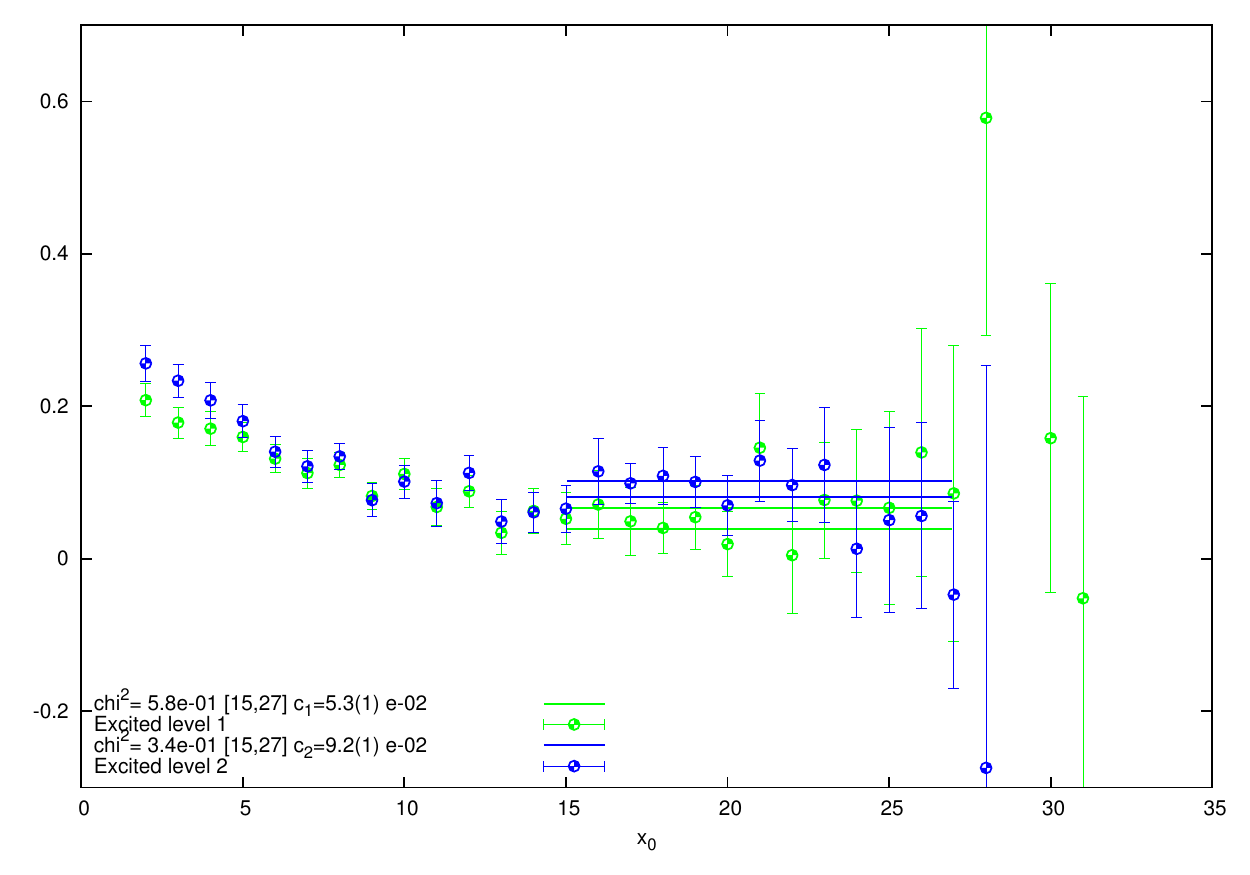}
\end{minipage}
\caption{In the left panel is shown the ground state effective mass of the $J^P=1^+$, $I=0$ channel, in lattice units $1.163(5)$.
In the right panel are shown the mass splittings for the first two excited states.}
\label{fig:iso0}
\end{center}
\end{figure}

\begin{figure}[htbp]
\begin{center}
\includegraphics[scale=0.9]{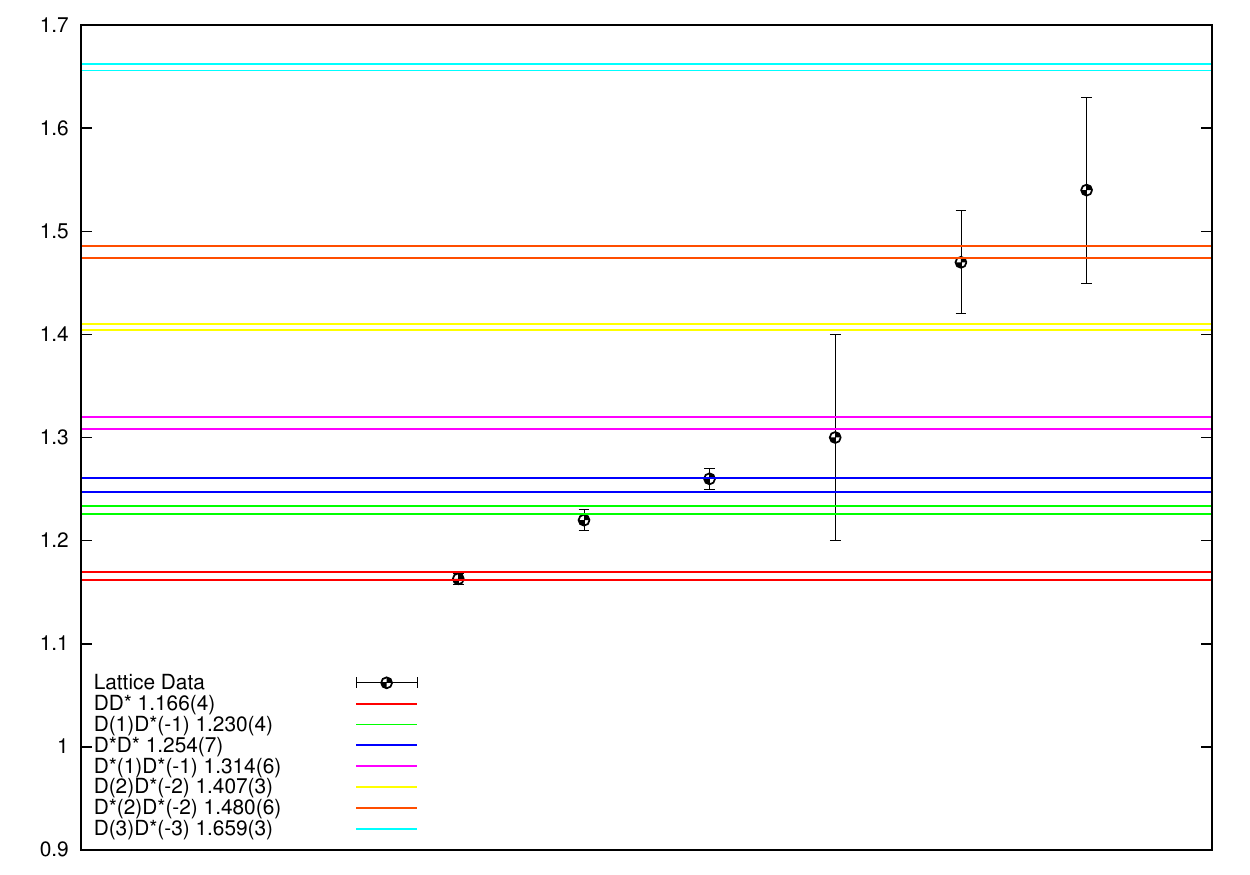}
\caption{Spectrum of the $J^P=1^+$, $I=0$ channel extracted from the basis of operators $\mathcal{O}_1\mathcal{O}_2\mathcal{O}_3$. Notice that we extract the state 
$D(1)D^*(-1)$, without inserting the corresponding operator. This is due to the presence of correlators 
constructed with point like inverted propagators.}
\label{fig:spect1}
\end{center}
\end{figure}

\begin{figure}[htbp]
\begin{center}
\includegraphics[scale=0.9]{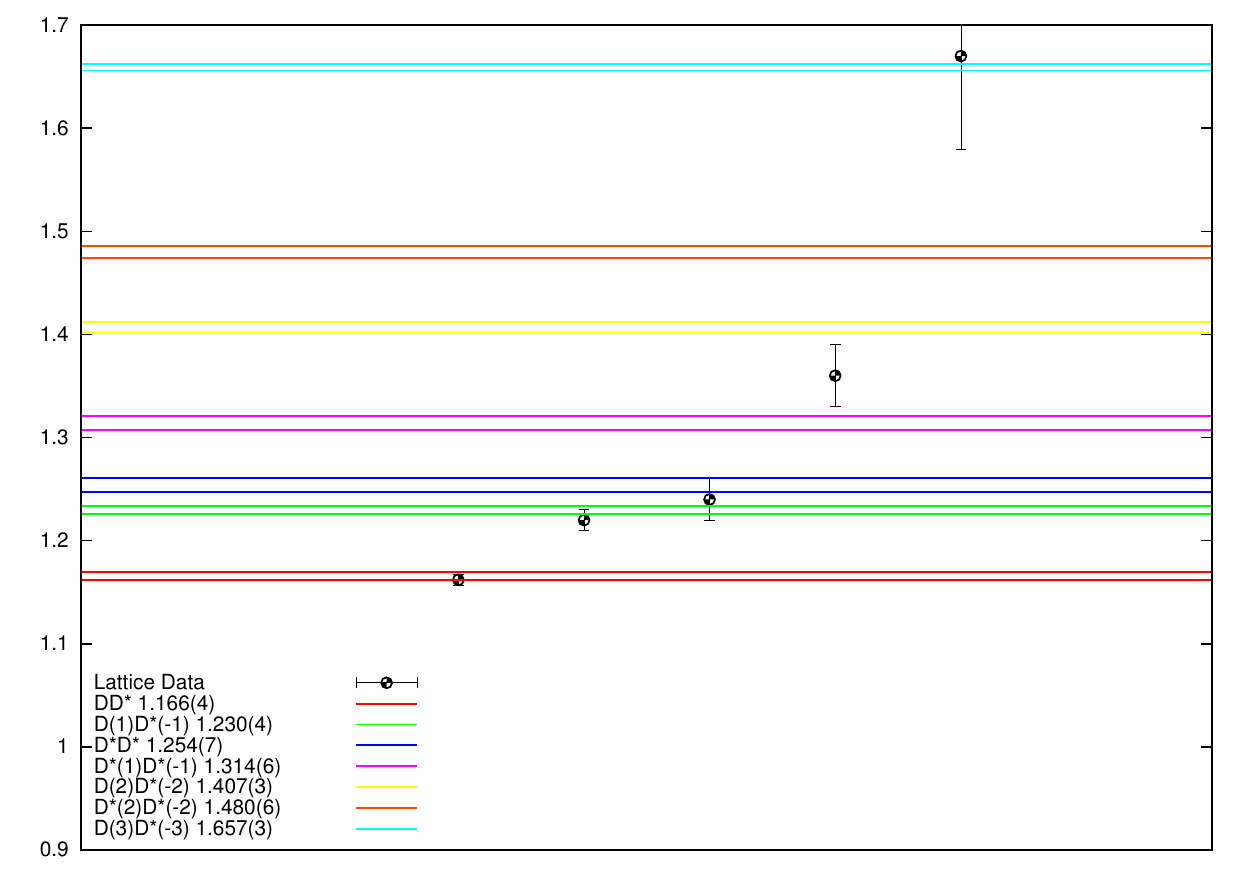}
\caption[Spectrum]{Spectrum of the $J^P=1^+$, $I=0$ channel extracted from the basis of operators $\mathcal{O}_2\mathcal{O}_3\mathcal{O}_4\mathcal{O}_5$.
 The less accuracy of the spectrum, in contrast with Fig.~\ref{fig:spect1} can be explained observing that in the off diagonal correlation functions between $D$ $D^*$ and $D^*$ $D^*$ operators, 
the spin structure makes them noisier than the other correlation functions of the GEP.}
\label{fig:spect2}
\end{center}
\end{figure}

\begin{figure}[htbp]
\begin{center}
\begin{minipage}[l]{.40\textwidth}
\centering
\includegraphics[scale=0.6]{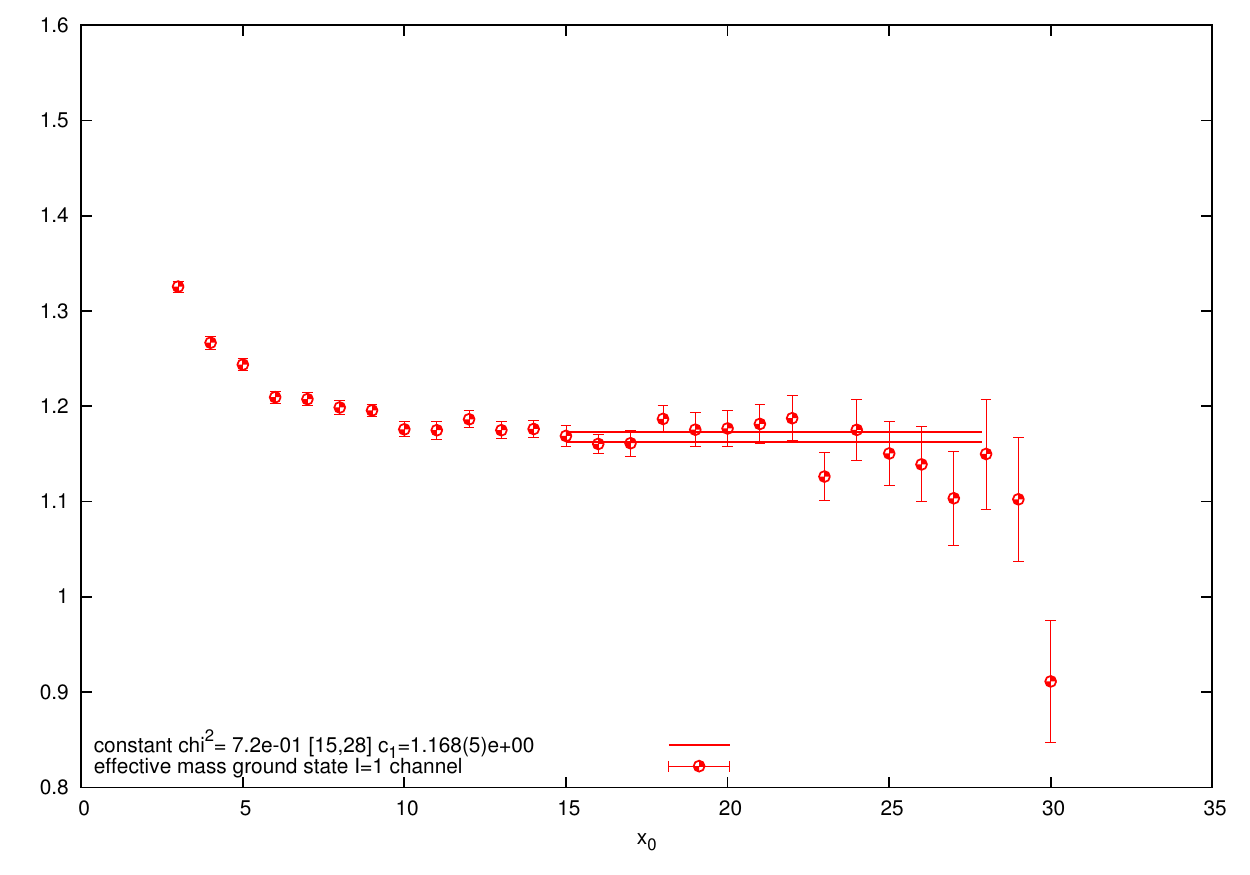}
\end{minipage}
\hspace{10mm}
\begin{minipage}[c]{.40\textwidth}
\centering
\includegraphics[scale=0.6]{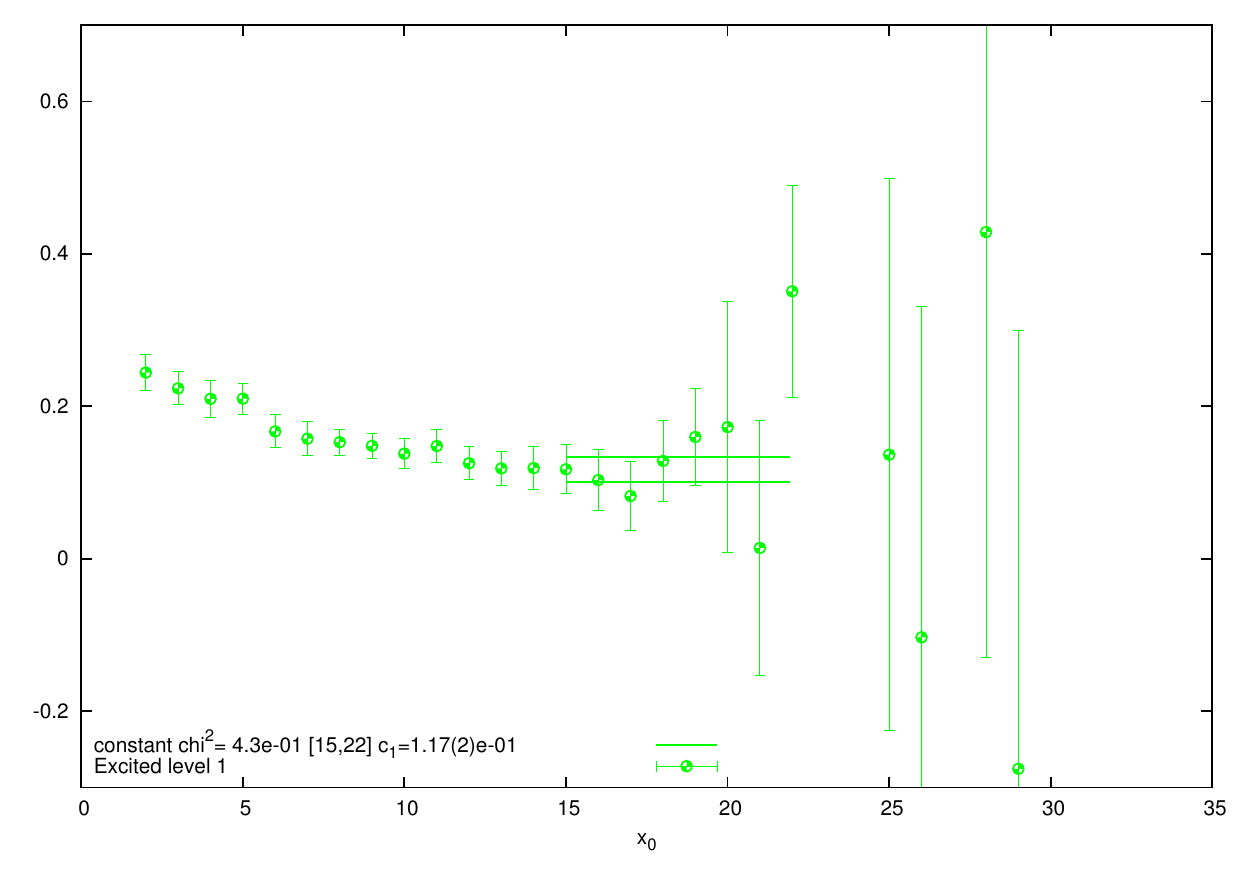}
\end{minipage}
\caption{n the left panel is shown the ground state effective mass of the $J^P=1^+$, $I=0$ channel, in lattice units $1.168(5)$.
In the right panel is shown the mass splitting of the first excited level.}
\label{fig:isospin1}
\end{center}
\end{figure}

We performed our analysis using $128$ CLS configurations with volume $L^3 \times T=32^3 \times 64$, $N_f=2$ number of 
sea flavors, non perturbatively $O(a)$ improved.
The bare gauge coupling is $\beta=5.2$, reproducing a lattice spacing of $a=0.075$ fm. Sea quarks have the same mass $\kappa_{sea}=0.13580$ corresponding to 
a pion mass, in physical units, of $m_{\pi} \sim 490$ MeV. For simulating the heavy quark charm we used a mass $\kappa_{charm}=0.13022$, lighter than the physical 
charm mass.
Preliminarily, we determined $D$, $D^*$ meson masses in our setup solving a $2\times 2$ GEP separately for the $J^{PC}=0^{-+}$ and $J^{PC}=1^{--}$ quantum numbers.
The two dimensional basis used is
\begin{equation}
\mathcal{O}(x)=\bar c(x) \Gamma l(x), \hspace{10mm} \widetilde{\mathcal{O}}=\bar c(x)  \hat S \, \Gamma \, \hat S l(x),
\end{equation}
with $\Gamma=\gamma_5$ for the $D$ channel and $\Gamma=\gamma_{\mu}$ for the $D^*$, $\hat S$ is the smearing operator.
As we need both point like and stochastic sources, we solved each GEP separately for both of them and performed the jackknife 
sum of the eigenvalues obtained with different sources.
The effective mass of the resulting eigenvalues is shown in Fig.~\ref{fig:soglie}. Using the extracted mass values, in lattice units, 
\begin{equation}
 am_D=5.389(8) \times 10^{-1} \hspace{10mm} am_{D^*}=6.27(3)\times 10^{-1}
\end{equation}
we determine the two particle thresholds for two non interacting mesons, see right panel of Fig.~\ref{fig:soglie}.
The aim of this preliminary analysis is that, in order to identify additional states, we need to know, at least approximatively for the non interacting case, 
the two meson state energies
that we expect to observe in the spectrum.
In the left panel of Fig.~\ref{fig:iso0}, we show the ground state in the heavy light sector with $I=0$ resulting from the GEP with input operators $\mathcal{O}_1,\mathcal{O}_2,
\mathcal{O}_3$.
Since the fluctuations of the higher eigenvalues are correlated with those of the ground state, it is convenient to determine only the mass splittings
\begin{equation}
\Delta M_i=-\ln \frac{\lambda_i(t) \lambda_0(t-1)}{\lambda_i(t-1)\lambda_0(t)},
\label{eq:splittings}
\end{equation}
with $i$ the index of the $i-$th eigenvalue, see Fig.\ref{fig:iso0} right panel.
The spectrum of the states, obtained adding the splittings of the excited levels to the ground state, is shown in Figs.~\ref{fig:spect1}, \ref{fig:spect2},
respectively for the basis $\mathcal{O}_1,\mathcal{O}_2,\mathcal{O}_3$ and the basis $\mathcal{O}_2,\mathcal{O}_3,\mathcal{O}_4,\mathcal{O}_5$.
Although we use a smaller basis of operators for the spectrum in Fig.~\ref{fig:spect1}, we observe that the numerical results are more accurate.
This can be due to off diagonal elements between $D$ $D^*$ and $D^*$ $D^*$ operators that, in virtue of the different spin structures, are numerically noisier than 
the other correlation functions in the GEP.
The spectrum of the channel $J^P=1^+$ with $I=1$ is noisier than the $I=0$ case and we cannot show the relative spectrum.
In Fig.~\ref{fig:isospin1} is reported the effective mass of the ground state and the mass splitting for the first excited level.
\section{Conclusions}
We set up a lattice study of QCD  states containing unambiguously four valence quarks.
The preliminary results obtained doesn't show any unknown energy level.
This doesn't exclude the possibility that a resonance could appear in one of the channel considered in this analysis.
In order to clarify the situation, it could be better to introduce a larger basis of operators.
On the other hand, the application of the Luscher method~\cite{Luscher:1991cf} to extract the scattering phase in the scattering channels considered is 
particularly challenging due to both 
the presence of a spin$-1$ particle and a very small elastic scattering window.
An interesting alternative could be the $D$ $D$ scattering channel. In that case, if a scalar resonance were found it could be identified as one of the 
bad scalar flavored tetraquarks.

\end{document}